\documentclass[aps,prl,twocolumn]{revtex4}
\usepackage{graphicx}
\usepackage{amsmath}
\usepackage{soul,color} 
\sethlcolor{red}

\begin{document}

\title{Spin Collective Modes of Two-Species Fermi Liquids: Helium-3 and Atomic Gases near the Feshbach Resonance}

\author{H P Dahal$^1$}
\author{S Gaudio$^2$}
\author{J D Feldmann$^1$}
\author{K S Bedell$^1$}

\affiliation{$^1$Department of Physics, Boston College, 140 Commonwealth Ave, Chestnut Hill, MA, 02467, USA}
\affiliation{$^2$Department of Physics, University of Rome La Sapienza, P.le Aldo Moro 2, 00185 Rome, Italy}

\begin{abstract}
We present theoretical findings on the spin collective modes of a two-species Fermi liquid, prepared alternatively in a polarized equilibrium or a polarized non-equilibrium state.  We explore the effects on these modes of a diverging s-wave scattering length, as occurs near a Feshbach resonance in a Fermionic atomic gas.  We compare these atomic gas modes with those of the conventional Helium-3 system, and we find that they differ from the conventional systems, and that the gap and spin stiffness are tunable via the Feshbach resonance.
\end{abstract}

\date{June 30, 2008}

\maketitle

While the BCS and BEC states in atomic gases garner wide interest across many fields \cite{Cornell,Ketterle,Stoof,Regal}, investigations into the normal state of atomic gases, i.e. above the superfluid phase transition, can also provide interesting results and important insights into the properties of these gases and other related systems.  For instance, theoretical studies directed towards the density excitations of atomic gases in the hydrodynamic, collisionless, and intermediate regimes \cite{Bruun,Pedri,Massignan} paved the way for experimental investigations into the excitation spectra \cite{Kinast1,Bartenstein} and the discovery of surprising features.  Application of Fermi liquid theory to density fluctuations of atomic gases has also led to interesting predictions and results \cite{Yip,Akdeniz}.

Using Fermi liquid theory (FLT) we study qualitatively the collective transverse spin modes, also known as Silin modes \cite{Silin}, of an atomic gas in the normal state near the Feshbach resonance (FBR), for a polarized equilibrium (PEQ) system and a polarized non-equilibrium (PNEQ) system (we define these systems below).  PEQ modes have been studied in other systems, most notably in Helium gases, but we report here that the specific application of the theory to atomic gases near an FBR provides unique possibilities for the tuning of PEQ and PNEQ modes and for the exploration of these modes across a wide range of interaction strengths.

A chief characteristic of the FBR is the divergence of the bare s-wave scattering length as the external magnetic field is tuned towards the resonance.  At low enough temperatures, only the s-wave scattering process is allowed, thus the characteristic scattering length is effectively determined by the s-wave scattering length alone.  Through the induced interaction model we calculate and plot the quantitative relationship between the s-wave scattering length and the Fermi liquid interaction parameters (LPÕs).  From this relationship we show the consequences of such a diverging scattering length on the spin modes of Fermi liquid theory.  We find that the gaps of particular modes are tunable near the FBR, as is the quadratic dependence (spin stiffness) on the wave vector.

The inhomogeneity of a confined atomic gas is not expected to affect the Fermi liquid results significantly \cite{Migdal}.  In lower dimensions, however, the confining potential restricts the motion of the atoms in certain directions, thus Fermi liquid results have been shown to change significantly for cigar-shaped traps \cite{Capuzzi}.  Therefore, all of the following calculations are done for the three-dimensional, homogeneous case.

We begin by defining the FLT parameters that are used throughout this paper.  For a complete reference, we refer the reader to the literature \cite{Pines,LFLT,Hari}, which we will follow closely in form and notation.  In FLT the variation of the energy $\delta E$ due to a variation of the quasi-particle (q-p) distribution function $\delta n_{\bf p\sigma}$ from its ground state can be written as

\begin{equation}
\label{eq1}
 \delta E= \frac{1}{V}\sum_{\textbf{p}\sigma}\varepsilon_{\textbf{p}\sigma}^0\delta n_{\textbf{p}\sigma} + \frac{1}{V^2}\frac{1}{2}\sum_{\textbf{p}\sigma,\textbf{p}'\sigma'}f_{\textbf{p}\sigma,\textbf{p}'\sigma'}\delta n_{\textbf{p}\sigma} \delta n_{\textbf{p}'\sigma'} + ...,
\end{equation}

\noindent where $\varepsilon_{\textbf{p}\sigma}^0$ is the single-particle excitation spectrum, and the q-p interaction energy $f$ is a second functional derivative of the total energy

\begin{equation}
f_{\textbf{p}\sigma,\textbf{p}'\sigma '}=V^2 \frac{\delta^2 E}{\delta n_{\textbf{p}\sigma} \delta n_{\textbf{p}'\sigma'}}
\end{equation}

\noindent which can be separated into symmetric and anti-symmetric parts, $f_{\bf pp'}=f_{\bf pp'}^s+f_{\bf pp'}^a{\bf \sigma\cdot\sigma '}$, where $\sigma$ denotes the spin state of the q-p, which are in turn related in a two-spin-component system to $f^{\uparrow\uparrow}$ and $f^{\uparrow\downarrow}$.  Furthermore, $f$ can be written in the usual way in terms of the Legendre expansion of the angle $\theta$ between ${\bf p}$ and ${\bf p'}$, $f_{\bf pp'}^{s,a}=\sum_{l=0}^{\infty}f_l^{s,a}P_{l}(\cos(\theta))$.  The dimensionless Landau parameters (LP) are obtained by the relation $F_l^{s,a}=N(0)f_l^{s,a}$, where $N(0)$ is the q-p density of states at the Fermi surface.  The definition of the spin polarization density $m$ in a two-component Fermi system is $m=\delta n_{\uparrow}-\delta n_{\downarrow}$.

A PEQ system is a spin-polarized system that has a net polarization that arises from, and is in equilibrium with, a polarizing external magnetic field.  Furthermore, the polarization is simply related to the external magnetic field strength and the LP's, and is given by $m_0=\delta n_{\uparrow}- \delta n_{\downarrow}=H(N(0)/(1+F_0^a))$, where $H$ is the magnitude of the applied external magnetic field, and $\hbar,\gamma \equiv 1$.

A PNEQ system is one in which the system is polarized, but is not in equilibrium with an external magnetic field.  The system is instead kept in the polarized state by external means other than a magnetic field, for instance, by constantly pumping a certain spin species into the system in order to maintain a finite polarization, or by using laser-induced transitions to convert one spin species to the other.  We will denote the PNEQ polarization density by $m'$.

With the parameters of the system suitably defined, we now briefly review the transverse spin collective mode calculation as derived in \cite{LFLT} and then discuss the behavior of the modes in Helium-3 and in an atomic gas near an FBR.  We begin with the familiar Landau kinetic equation (LKE), which governs the temporal and spatial evolution of a local spin polarization density,

\begin{eqnarray}
\frac{\partial {\bf m}_{\bf p}}{\partial t}&+&\frac{\partial}{\partial r_i}(\frac{\partial \epsilon_{\bf p}}{\partial p_i}{\bf m}_{\bf p} +\frac{\partial {\bf h}_{\bf p}}{\partial p_i}n_{\bf p})-\frac{\partial}{\partial p_i}(\frac{\partial \epsilon_{\bf p}}{\partial r_i}{\bf m}_{\bf p}+ \nonumber\\
&&\frac{\partial {\bf h}_{\bf p}}{\partial r_i}n_{\bf p})=(\frac{\partial {\bf m}_{\bf p}}{\partial t})_{prec.}+(\frac{\partial {\bf m}_{\bf p}}{\partial t})_{coll.}
\label{LKE}
\end{eqnarray}

\noindent where $m_{\bf p}\equiv m_{\bf p}({\bf r},t)=\frac{1}{2}\sum_{\alpha \alpha '}{\bf \tau}_{\alpha \alpha '}[n_{\bf p}({\bf r},t)]_{\alpha '\alpha}$ is the local spin polarization, and ${\bf h_p}=-\frac{1}{2}H_0+2\int\frac{d^3p'}{(2\pi)^3}f_{\bf pp'}^a{\bf \sigma_{p'}}$ is the effective internal field.  Eq. (\ref{LKE}) describes the evolution of a spin perturbation in an interacting Fermi system.  From this equation the expression for the evolution of a transverse spin perturbation can be derived.  The right hand side contains two terms, the precession term and the collision term.  For the case of cold atomic gases we assume that the collision term can be taken to be zero, and we retain the precession term.  A solution to this equation is achieved through a Fourier transform and spherical harmonic expansion of the Fermi surface deformation.  We truncate the harmonic expansion at $l=1$, since the $l=2$ term gives a small correction and does not change the structure of the solution in the limit $q \rightarrow 0$ \cite{Bedell2,Hari}.  From \cite{Hari}, the two PNEQ solutions are

\begin{equation}
\omega_{0,PNEQ}^{\pm}= \pm \frac{\frac{1}{3}(1+F_0^a)(1+\frac{F_1^a}{3})(qv_F)^2}{\frac{2m'}{N(0)}(F_0^a-\frac{F_1^a}{3})}
\label{omega0PNEQ}
\end{equation}

\begin{eqnarray}
\omega_{1,PNEQ}^{\pm}&=& \mp \frac{2m'}{N(0)} (F_0^a-\frac{F_1^a}{3})) \nonumber \\
&&(1+\frac{\frac{1}{3}(1+F_0^a)(1+\frac{F_1^a}{3})(qv_F)^2}{(\frac{2m'}{N(0)}(F_0^a-\frac{F_1^a}{3}))^2})
\label{omega1PNEQ}
\end{eqnarray}

\noindent The PEQ dispersion is given by the addition of the Larmor frequency $\omega_L$ to the PNEQ results (note that an accompanying change of notation $m'\rightarrow m_0$ is also required).

\begin{equation}
\omega_{0,PEQ}^{\pm}=\pm \omega_L \pm \frac{\frac{1}{3}(1+F_0^a)(1+\frac{F_1^a}{3})(qv_F)^2}{\frac{2m_0}{N(0)}(F_0^a-\frac{F_1^a}{3})}
\label{omega0PEQ}
\end{equation}

\begin{eqnarray}
\omega_{1,PEQ}^{\pm} &=&\pm \omega_L \mp \frac{2m_0}{N(0)} (F_0^a-\frac{F_1^a}{3})) \nonumber \\
&&(1+\frac{\frac{1}{3}(1+F_0^a)(1+\frac{F_1^a}{3})(qv_F)^2}{(\frac{2m_0}{N(0)}(F_0^a-\frac{F_1^a}{3}))^2})
\label{omega1PEQ}
\end{eqnarray}

The p-h continuum dispersion is given by the relation

\begin{equation}
\omega_{ph,PNEQ}^{\pm}=\pm \frac{2m'F_0^a}{N(0)}+{\bf q}\cdot {\bf v}
\label{phcont1}
\end{equation}

\begin{equation}
\omega_{ph,PEQ}^{\pm}= \pm \omega_L \pm \frac{2m_0F_0^a}{N(0)}+{\bf q}\cdot {\bf v}
\label{phcont2}
\end{equation}

\begin{figure}
\includegraphics[width=3.3 in,angle=0]{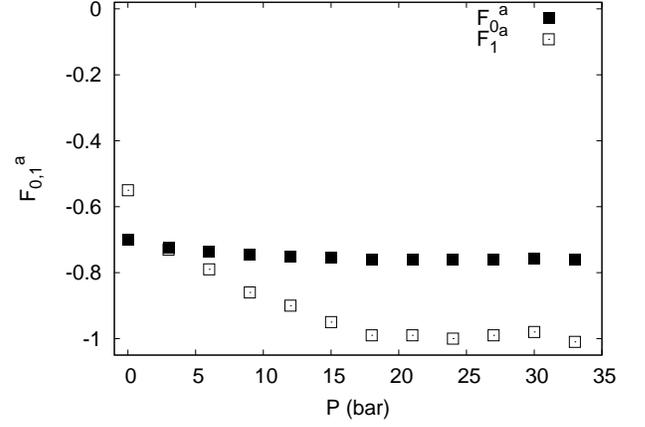}
\caption{Landau parameters in the Helium-3 system at varying pressures (taken from Greywall (1983) \cite{Greywall}.  Notice the small change in absolute value $F_0^a$ as compared with the large change in $F_0^a$ in an atomic gas near a Feshbach resonance (see Fig. \ref{LP}).  We also note here that in He-3, $F_0^a$ and $F_1^a$ have the same sign, whereas we have calculated that they have difference signs in an atomic gas.}
\label{LPHe3}
\end{figure}

It is instructive to apply these solutions first to the He-3 system, since this system has been studied extensively (e.g. see \cite{LFLT} and references therein).  In He-3, the anti-symmetric Landau parameters at zero pressure are $F_0^a=-0.7$, $F_1^a=-0.55$ \cite{Greywall}, and the LP's for different pressures are shown in Fig. \ref{LPHe3}.  As seen in the figure, large changes in pressure in the He-3 system lead to only small variation in the anti-symmetric LP's (it is the symmetric LP's that vary greatly with pressure in He-3).

\begin{figure}
\includegraphics[width=3.3 in,angle=0]{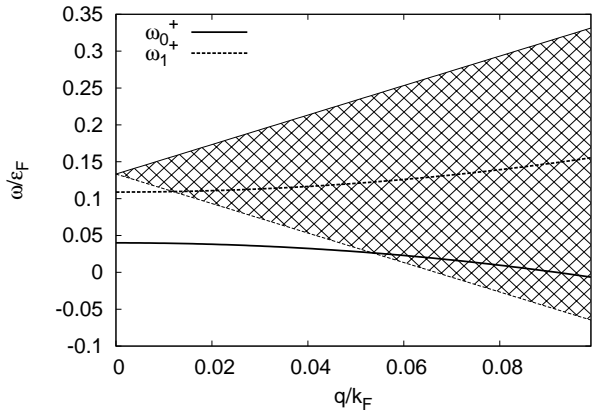}
\caption{The PEQ spin collective mode dispersion (theory) for Helium-3 at a pressure of 3 bar.  The non-zero $F_1^a$ brings the current mode out of the particle-hole continuum for very small values of $q$.}
\label{HeDisp}
\end{figure}

Theoretical calculations for the transverse spin wave dispersion for Helium-3 at $P=3$ bar is shown in Fig.\ \ref{HeDisp}.  These modes were theoretically predicted by Abrikosov and Dzyaloshinski \cite{AbrikDzyal} in 1959, and the modes have been experimentally observed in $^3$He and $^3$He-$^4$He mixtures \cite{Corruccini,Owers-Bradley,Masuhara,Ishimoto}.

\begin{figure}
\includegraphics[width=3.3 in,angle=0]{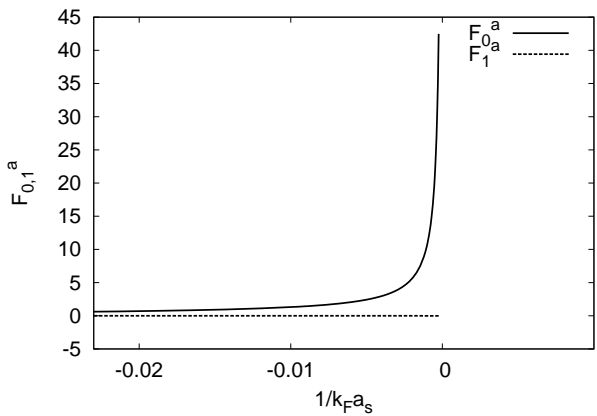}
\caption{Calculated Landau parameters, $F_0^a$ and $F_1^a$, for a gas of $^6$Li atoms in the appropriate high-field seeking spin states near the 834-Gauss Feshbach resonance.  The horizontal axis is the inverse of the bare scattering length $a_s$ times the fermi wave vector $k_F$.}
\label{LP}
\end{figure}

In order to evaluate the collective mode solutions for an atomic gas near an FBR, values for the pertinent LP's must be calculated.  We calculate the LP's using the induced interaction model \cite{Babu,Quader}, which provides a formal relation between the scattering length and the LP's, and our results are shown in Fig.\ \ref{LP}.  The figure shows the parameter $F_0^a$ diverging towards $+\infty$ as the Feshbach resonance is approached from the attractive side ($a_s<0$).  The parameter $F_1^a$ is less than zero and remains small in magnitude near the resonance.

\begin{figure}
\includegraphics[width=3.3 in,angle=0]{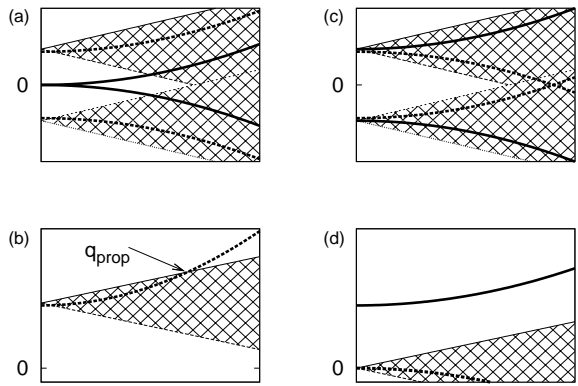}
\caption{The qualitative dispersion ($\omega$ vs. $q$) behavior of the atomic gas (a) precessional mode $\omega_{0,PNEQ}^{\pm}(q)$ and current mode $\omega_{1,PNEQ}^{\pm}(q)$ far from the FBR, i.e. in a weakly interacting Fermionic atomic gas; (b) current mode $\omega_{1,PNEQ}^-(q)$. The label $q_{prop}$ indicates the wave vector value at which the mode exits the particle-hole continuum, which varies directly with the non-equilibrium magnetization $m'$; (c) precessional mode $\omega_{0,PEQ}^{\pm}(q)$ and current mode $\omega_{1,PEQ}^{\pm}(q)$ far from the FBR; (d) precessional mode $\omega_{0,PEQ}^+(q)$ and current mode $\omega_{1,PEQ}^+(q)$ near the FBR, where the spin current mode lies beneath the spin precessional mode, and is nearly gapless.  In all plots, the solid line indicates the precessional mode, the dashed line indicates the current mode, and the shaded region indicates the particle-hole continuum.}
\label{fig4}
\end{figure}

With the behavior of the LP's near the FBR determined, the dispersion relations can be evaluated for the atomic gas system. The spin modes in the PNEQ system are characterized by a gapless spin precessional mode and a gapped spin current mode.  The gap in the current mode is given by $\omega_{1,PNEQ}^+(q=0)=2m'F_0^a/N(0)-2m'F_1^a/3N(0)$.  The first part arises from the effective internal field produced by the q-p's, equal to $2m'F_0^a/N(0)$, and the second part from the modification of the field due to the Fermi surface distortion, equal to $-2m'F_1^a/3N(0)$ \cite{LFLT}.  The qualitative behavior of the PNEQ modes far from the FBR in an atomic gas is shown in Fig. \ref{fig4} (a).  The current mode gap could be tuned by manipulating the polarization $m'$ of the system.  For instance, increased injection of spin ''up'' $^6$Li atoms ($F=1/2$,$m_F=+1/2$) would result in an increase in the gap of the current mode.

The spin stiffness of the PNEQ modes can be seen to be inversely proportional to the polarization density $m'$.  Thus for small polarization, the spin stiffness is very large.  Therefore, the emergence of the current mode from the p-h continuum, as indicated in Fig.\ \ref{fig4} (b)  by $q_{prop}$, could be tuned to occur at low values of the wave number $q$ by manipulation of the polarization density.

In Fig.\ \ref{fig4} (c) we plot the qualitative behavior of the PEQ modes far from the FBR.  The gap of the PEQ spin precessional mode is given simply by the Larmor frequency.  However, the gap of the PEQ spin current mode is inversely proportional to $F_0^a$, as can be seen in the expression for the gap after substitution for the equilibrium value of the polarization, $\omega_{1,PEQ}^+(q=0)=+\omega_L\frac{1+F_1^a/3}{1+F_0^a}$.  Thus if our calculations for the LP's are correct, this mode could be made nearly gapless near an FBR due to the divergence of $F_0^a$, and thus be brought below the spin precessional mode.  This scenario is shown in Fig.\ \ref{fig4} (d).

Due to the equilibrium relation $m_0=H_0N(0)/(1+F_0^a)$, the spin stiffness in a PEQ system behaves differently than in a PNEQ. In the limit of $F_0^a \gg 1$, the spin stiffness $D$ is approximately proportional to $D\sim 1/m_0$, and $m_0\sim 1/F_0^a$.  Thus the spin stiffness increases linearly with increasing $F_0^a$.  This would mean that as the FBR is approached from the attractive side, the spin stiffness would increase, and the current mode would exit the p-h continuum at lower and lower values of $q$.

These collective spin mode effects are unique to the atomic gas system near an FBR above the superfluid transition.  No other Fermi liquid system offers such easily tunable effects.  An experimental investigation into these collective mode effects could yield the observation of a variety of new physical phenomena, and lead to a better understanding of collective mode phenomena in other systems and temperature regimes, as well.

In conclusion, we have presented our findings on the collective spin modes of a three-dimensional, homogeneous Fermionic atomic gas in the normal phase.  We have discussed the general mode behavior for PEQ and PNEQ systems, as well as specific behavior for $^3$He and atomic gases.  In contrast to $^3$He, we have shown that the gap and spin stiffness of the atomic gas modes can be tuned near the Feshbach resonance by manipulation of the s-wave scattering length.  We postulate that these modes and effects could be experimentally detected and confirmed in atomic gases, and thus lead to a better understanding of dilute atomic gases near the Feshbach resonance, as well as the collective spin modes of a variety of Fermi liquid systems.  We are currently investigating further theoretical implications of the results presented above, including the thermodynamic repercussions of coupling the density and spin excitations.

This work has been supported by a grant from the Rourke Professor of Physics.


\begin{thebibliography}{1}

\bibitem{Cornell} 
M. H. Anderson, J. R. Ensher, M. R. Mathews, C. E. Wieman, and E. A. Cornell, Science {\bf 269}, 198 (1995)

\bibitem{Ketterle} 
K. B. Davis, M. O. Mewes, M. R. Andrews, N. J. van Druten, D. S. Durfee, D. M. Kurn, and W. Ketterle, Phys. Rev. Lett. {\bf 75}, 3969 (1995)

\bibitem{Stoof}
H. T. C. Stoof, M. Houbiers, C. A. Sackett, and R. G. Hulet, Phys. Rev. Lett. {\bf 76}, 10 (1996)

\bibitem{Regal}
C. A. Regal, M. Greiner, and D. S. Jin, Phys. Rev. Lett. {\bf 92}, 083201 (2004)

\bibitem{Bruun}
G. M. Bruun and C. W. Clark, Phys. Rev. Lett. {\bf 83}, 5415 (1999) 

\bibitem{Pedri}
P. Pedri, D. Guery-Odelin, and S. Stringari, Phys. Rev. A {\bf 68}, 043608 (2003)

\bibitem{Massignan}
P. Massignan, G. M. Bruun, and H. Smith, Phys. Rev. A {\bf 71}, 033607 (2005)

\bibitem{Kinast1}
J. Kinast, S. L. Hemmer, M. E. Gehm, A. Turlapov, and J. E. Thomas, Phys. Rev. Lett. {\bf 92}, 150402 (2004)

\bibitem{Bartenstein}
M. Bartenstein, A. Altmeyer, S. Riedl, S. Jochim, C. Chin, J. H. Denschlag, and R. Grimm, Phys. Rev. Lett. {\bf 92}, 203201 (2004)

\bibitem{Yip}
S.-K. Yip and T.-L. Ho, Phys. Rev. A {\bf 59}, 4653 (1999)

\bibitem{Akdeniz}
Z. Akdeniz, P. Vignolo, and M. P. Tosi, Phys. Lett. A {\bf 311}, 246 (2003)

\bibitem{Silin}
V. P. Silin, J. Exptl. Theoret. Phys. {\bf 33}, 495 (1957) [Sov. Phys.-JETP {\bf 6}, 387 (1957)]

\bibitem{Migdal}
A. B. Migdal, {\it Nuclear Theory:  The Quasiparticle Method}, W. A. Benjamin, Inc., (1968)

\bibitem{Capuzzi}
P. Capuzzi, P. Vignolo, F. Federici, and M. P. Tosi, J. Phys. B: At. Mol. Opt. Phys. {\bf 39}, S25 (2006)

\bibitem{Mihaila}
B. Mihaila, S. A. Crooker, K. B. Blagoev, D. G. Rickel, P. B. Littlewood, and D. L. Smith, cond-mat/0601011v1 (January 2006)

\bibitem{LeggettRice}
A. J. Leggett and M. J. Rice, Phys. Rev. Lett. {\bf 20}, 586 (1968)

\bibitem{Pines} 
Pines and Nozieres, Phys. Rev. {\bf 60}, 100 (1958)

\bibitem{LFLT} 
G. Baym and C. Pethick, {\it Landau Fermi-Liquid Theory: Concepts and Applications}, John Wiley and Sons, Inc., (1991)

\bibitem{Hari}
K. S. Bedell and H. P. Dahal, Phys. Rev. Lett. {\bf 97}, 047204 (2006)

\bibitem{Bedell2}
K. S. Bedell, D. E. Meltzer, Phys. Rev. B {\bf 33}, 4543 (1986)

\bibitem{Greywall}
D. S. Greywall, Phys. Rev. B {\bf 27}, 2747 (1983)

\bibitem{AbrikDzyal}
A. A. Abrikosov and I. E. Dzyaloshinski, Sov. Phys. JETP {\bf 8}, 535 (1959)

\bibitem{Corruccini}
L. R. Corruccini, D. D. Osheroff, D. M. Lee and R. C. Richardson, J. Low Temp. Phys {\bf 8}, 229 (1972)

\bibitem{Owers-Bradley}
J. R. Owers-Bradley, H. Chocholacs, R. M. Mueller, Ch. Buchal, M. Kubota and F. Pobell, Phys. Rev. Lett. {\bf 51}, 2120 (1983)

\bibitem{Masuhara}
N. Masuhara, D. Candela, D. O. Edwards, R. F. Hoyt, H. N. Scholz, D. S. Sherrill and R. Combescot, Phys. Rev. Lett. {\bf 53}, 1168 (1984); D. Candela, D. O. Edwards, A. Heff, N. Masuhara, Y. Oda and D. S. Sherrill, Phys. Rev. Lett. {\bf 61}, 420 (1988)

\bibitem{Ishimoto}
H. Ishimoto, H. Fukuyama, N. Nishida, Y. Miura, Y. Takano, T. Fukuda, T. Tazaki and S. Ogawa, Phys. Rev. Lett. {\bf 59}, 904 (1987)

\bibitem{Babu}
S. Babu and G. E. Brown, Annals of Physics {\bf 78}, 1 (1973)

\bibitem{Quader}
K. F. Quader and K. S. Bedell, J. Low Temp. Phys {\bf 58}, 89 (1985)

\bibitem{Haritobe}
H. Dahal, et al., (to be published)

\end{thebibliography}
\end{document}